\documentclass[11pt]{article}
\usepackage{amsmath,amssymb,color,slashed}

\usepackage{amsfonts}

\newcommand{\hoch}[1]{$\, ^{#1}$}

\newcommand{\be}{\begin{equation}}
\newcommand{\ee}{\end{equation}}
\newcommand{\bea}{\setlength\arraycolsep{2pt} \begin{eqnarray}}
\newcommand{\eea}{\end{eqnarray}}

\def\ft#1#2{{\textstyle{\frac{\scriptstyle #1}{\scriptstyle #2} } }}
\def\fft#1#2{{\frac{#1}{#2}}}

\def\0{{\sst{(0)}}}
\def\1{{\sst{(1)}}}
\def\2{{\sst{(2)}}}
\def\3{{\sst{(3)}}}
\def\4{{\sst{(4)}}}
\def\5{{\sst{(5)}}}
\def\6{{\sst{(6)}}}
\def\7{{\sst{(7)}}}
\def\8{{\sst{(8)}}}
\def\sst#1{{\scriptscriptstyle #1}}

\begin{document}

\begin{flushright}
\hfill{CAS-KITPC/ITP-275 \ \ \ \ KIAS-P11041}
\end{flushright}

\vspace{25pt}
\begin{center}
{\large {\bf Gauged Kaluza-Klein AdS Pseudo-supergravity }}

\vspace{10pt}

Haishan Liu\hoch{1}, H. L\"u\hoch{2,3} and Zhao-Long Wang\hoch{4}

\vspace{10pt}

\hoch{1}{\it Zheijiang Institute of Modern Physics\\
Department of Physics, Zheijiang University, Hangzhou 310027, China}

\vspace{10pt}

\hoch{2}{\it China Economics and Management Academy\\
Central University of Finance and Economics, Beijing 100081, China}

\vspace{10pt}

\hoch{3}{\it Institute for Advanced Study, Shenzhen University\\
Nanhai Ave 3688, Shenzhen 518060, China}

\vspace{10pt}

\hoch{4} {\it School of Physics, Korea Institute for Advanced Study,
Seoul 130-722, Korea}

\vspace{40pt}

\underline{ABSTRACT}
\end{center}

We obtain the pseudo-supergravity extension of the $D$-dimensional
Kaluza-Klein theory, which is the circle reduction of pure gravity
in $D+1$ dimensions.  The fermionic partners are pseudo-gravitino and pseudo-dilatino.  The full Lagrangian is invariant under the
pseudo-supersymmetric transformation, up to quadratic order in
fermion fields.  We find that the theory possesses a $U(1)$ global symmetry that can be gauged so that all
the fermions are charged under the Kaluza-Klein vector.  The gauging
process generates a scalar potential that has a maximum, leading to
the AdS vacuum.  Whist the highest dimension for gauged AdS
supergravity is seven, our gauged AdS pseudo-supergravities can
exist in arbitrary dimensions.

\vspace{15pt}

\thispagestyle{empty}





\newpage

\section{Introduction}

There are usually two criteria for a successful construction of
supergravities.  The first is that the degrees of freedom of the
bosonic  and fermionic fields should match.  This condition itself does
not provide much restriction on the construction. The second, which
is much more non-trivial, is that the bosonic part of the Lagrangian
should admit consistent Killing spinor equations, whose projected
integrability conditions give rise to the full set of the bosonic
equations of motion.  Let us consider eleven-dimensional
supergravity \cite{cjs} as an example.  The on-shell degrees of
freedom of the graviton and 3-form gauge potential match with those
of gravitino.  However, in addition to the Einstein-Hilbert action
and the kinetic term for the 3-form potential, eleven-dimensional
supergravity requires a Chern-Simons term for the 3-form with a
specific coupling. This term does not affect the total degrees of
freedom, but is essential for the consistency of the projected
integrability condition for the Killing spinor equations \cite{fog,gp,lw0}.

Recently, it was discovered that the low-energy effective action of
the bosonic string, which is an intrinsically non-supersymmetric
theory, admits consistent Killing spinor equations \cite{lpw}.  The
results were extended to include Yang-Mills fields and $\alpha'$
order corrections \cite{lw}, as well as the conformal anomaly term
\cite{llw}.  Based on these results, the pseudo-supergravity
extension of the bosonic string was constructed in \cite{lpwpsugr}.
Pseudo-supersymmetric partners, namely the pseudo-gravitino and
pseudo-dilatino, are introduced.  The full Lagrangian is invariant
under the pseudo-supersymmetric transformation, up to the quadratic
fermion order. The pseudo-supersymmetric theory can be extended by
coupling it to a Yang-Mills pseudo-supermultiplet.  This also allows
one to construct ``$\alpha'$ corrections'' involving quadratic
curvature terms.  An exponential dilaton potential term, associated
with the conformal anomaly for a bosonic string outside its critical
dimension, can also be pseudo-supersymmetrized.  Of course, in
$D=10$, where the degrees of freedom for the bosons and fermions
match, the theory may become fully supersymmetric after adding
quartic fermion terms.  The full ten-dimensional ${\cal N}=1$
supergravity with Yang-Mills supermultiplets were given in
\cite{Bvn,berderoo,berderoo2}.  However, when $\alpha'$ correction
terms are involved, the supersymmetry was proved only at the
quadratic fermion order \cite{berderoo,berderoo2}.

   Killing spinor equations for the $D$-dimensional Kaluza-Klein
theory that is the circle reduction of pure gravity in $D+1$
dimensions were obtained in \cite{lpw}. It was shown in \cite{llw}
that the Killing spinors can be charged under the Kaluza-Klein
vector.  This charging process generates a scalar potential, yielding
the full bosonic Lagrangian
\begin{equation}
e^{-1} {\cal L}= R - \ft12(\partial \phi)^2 - \ft14 e^{a\phi} F_\2^2
+g^2 (D-1) \Big( (D-3) e^{-\sqrt{\fft{2}{(D-1)(D-2)}}\,\phi} +
e^{\fft{\sqrt2\,(D-3)}{\sqrt{(D-1)(D-2)}}\,\phi}\Big)\,,\label{fullbosonlag}
\end{equation}
where $e=\sqrt{-g}$, $F_\2=dA_\1$ is the 2-form field strength for
the Kaluza-Klein vector $A_\1$, and
\begin{equation}
a=\sqrt{\fft{2(D-1)}{D-2}}\,.\label{avalue}
\end{equation}
In this paper, we obtain the gauged pseudo-supergravity extension
for this bosonic Lagrangian.

     The paper is organized as follows.  In section 2, we construct
ungauged Kaluza-Klein pseudo-supergravity that is
pseudo-supersymmetrization of the Kaluza-Klein theory. In
section 3, we pseudo-supersymmetrizing the scalar/gravity system
as a warmup exercise, since as we explained earlier, the gauging
process will generate a scalar potential.  In section 4, we combine
the results of sections 2 and 3, and obtain ungauged Kaluza-Klein
pseudo-supergravity with a scalar potential. We show that the scalar
potential has a single exponential term which has no fixed point.  The theory possesses $U(1)\sim SO(2)$ global symmetry that rotates the fermion fields. In section 5, we gauge the theory by letting all the fermions charged under the Kaluza-Klein vector. This effectively turns the $U(1)$ symmetry to become a local one. The gauging process extends the
previous scalar potential to involve two exponential terms. The new
scalar potential has a maximum, implying that the AdS spacetime is
its vacuum solution.  We conclude the paper in section 6.  We
summarize our results in the appendix. Since we
consider pseudo-supergravities in all dimensions, we also present the
fermion conventions in diverse dimensions in the appendix.

\section{Kaluza-Klein pseudo-supergravity}

The $D$-dimensional Kaluza-Klein theory is the $S^1$ reduction of
pure gravity in $D+1$ dimensions. The Lagrangian is given by
(\ref{fullbosonlag}) with $g=0$, namely
\begin{equation}
e^{-1}{\cal L}_{{\rm KK}, B} = R - \ft12(\partial \phi)^2 - \ft14
e^{a\phi} F^2_\2\,.
\end{equation}
The consistent Killing spinor equations for this were obtained in
\cite{lpw}.  Here we examine whether the theory can be
pseudo-supersymmetrized. We introduce pseudo-gravitino and dilatino
fields $(\psi_\mu^i, \lambda^i)$, and propose that the fermionic extension is given by
\begin{eqnarray}
e^{-1}{\cal L}_{{\rm KK}, F} &=& s^{ij}\Big[\ft12 \bar \psi^i_{\mu}
\Gamma^{\mu\nu\rho} D_\nu \psi^j_{\rho} + \ft12 \bar \lambda^i
{\slashed D} \lambda^j+ e_1\, \bar \psi^i_\mu \Gamma^\nu \Gamma^\mu
\lambda^j
\partial_\nu \phi\Big]\label{KKF}\\ 
&&+t^{ij}\Big[e_2\, \bar\psi^i_\mu \Gamma^{\mu\nu\rho\sigma}
\psi^j_\sigma + e_3\, \bar\psi^{i\nu}\psi^{j\rho} + e_4\, \bar
\psi^i_\mu \Gamma^{\nu\rho} \Gamma^\mu\lambda^j + e_5\, \bar
\lambda^i \Gamma^{\nu\rho}\lambda^j\Big] e^{\fft12a\phi}
F_{\nu\rho}\,,\nonumber
\end{eqnarray}
where the constant coefficients $e_1, \ldots, e_5$ are to be
determined.  Note that this is very different from the previous
supergravity construction, which is typically on a
specific dimension with specific type of fermion fields.  Here we
construct the theory in a generic dimension.  It is thus necessary
to summarize the properties of fermions in general dimensions.  We
adopt exactly the same convention given in \cite{lpwpsugr}, which
follows the convention of \cite{vanp}.  We present the convention in
Table 1 at the end of the appendix, and refer readers to
\cite{lpwpsugr,vanp} for details. In addition to the $\Gamma$-matrix
symmetries and spinor representations in diverse dimensions, we
also present the $s^{ij}$ and $t^{ij}$ (and also $u^{ij}$ that will
appear in later constructions) in Table 1.  The indices $i,j$ takes two
values, 1 and 2. From Table 1, we note that $s^{ij}$ and $t^{ij}$
can be either Kronecker $\delta^{ij}$ or $\varepsilon^{ij}$, where
$\varepsilon^{ij}=-\varepsilon^{ji}$ with $\varepsilon^{12}=1$.  In
dimensions where $s^{ij}=\delta^{ij}$, the fermions are two copies
of Majorana; when $s^{ij}=\varepsilon^{ij}$, they are of a single symplectic Majorana.  When $s^{ij}=t^{ij}$, which occurs only for $\beta=+1$, the fermions in (\ref{KKF}) can be
reduced to either a single copy of Majorana or symplectic Majorana,
depending on the dimensions.

Having presented the convention for the fermions in general
dimensions, we now give the ansatz for the pseudo-supersymmetric
transformation rules
\begin{eqnarray}
\delta\psi_\mu^i &=& D_\mu \epsilon^i  + \ft{{\rm i}}{8(D-2)}\, c_1
t^{ij}s^{kj}\Big(\Gamma_\mu \Gamma^{\nu\rho} -2(D-2)
\delta_{\mu}^{\nu} \Gamma^{\rho}\Big) e^{\fft12 a\phi} F_{\nu\rho}
\epsilon^k\,,\cr 
\delta \lambda^i &=& c_2 \Big[\Gamma^\mu \partial_\mu\phi\epsilon^i
+ \ft{{\rm i}}{4} c_3\,a\, t^{ij}s^{kj}\,\Gamma^{\mu\nu}
e^{\fft12a\phi}F_{\mu\nu} \epsilon^k\Big]\,,\cr 
\delta e^a_\mu &=& \ft14 s^{ij}\,\bar \psi_\mu^i \Gamma^a
\epsilon^j\,,\qquad \hbox{so}\qquad \delta g_{\mu\nu} = \ft12
s^{ij}\,\bar \psi^i_{(\mu} \Gamma_{\nu)} \epsilon^j\,,\cr 
\delta\phi &=& c_4\, s^{ij}\, \bar \lambda^i \epsilon^j\,,\cr 
\delta A_\mu &=& e^{-\fft12a\phi}\, t^{ij}\Big[c_5\, \bar\psi_\mu^i \epsilon^j
 + c_6\, \bar \lambda^i \Gamma_\mu \epsilon^j
\Big]\,,\label{kktrans}
\end{eqnarray}
where the coefficients $c_1,\ldots, c_6$ are to be determined.  Note
that the pseudo-supersymmetric transformation rules for the
fermionic fields $(\psi^i_\mu,\lambda^i)$ are inspired by the Killing
spinor equations obtained in \cite{llw} for single copy of fermions.

        We now require that the full Lagrangian
\begin{equation}
{\cal L}_{\rm KK} = {\cal L}_{{\rm KK}, B} + {\cal L}_{{\rm KK}, F}
\end{equation}
be invariant by the pseudo-supersymmetric transformation
(\ref{kktrans}), up to quadratic order in fermion fields.  We find
that this fixes the coefficients in the ansatze completely, given by
\begin{eqnarray}
&&e_1 = \fft{\rm i}{2\sqrt{2\beta}}\,,\quad e_2=\fft{{\rm
i}\sqrt{\beta}}{16}\,,\quad e_3=\fft{{\rm i}\sqrt{\beta}}8\,,\quad
e_4 = \fft{a}{8\sqrt{2\beta}}\,,\quad e_5=- \fft{ {\rm i}\,
D\sqrt{\beta}}{16(D-2)}\,,\cr 
&& c_1=c_3=\sqrt{\beta}\,,\quad c_2=\fft{{\rm
i}\sqrt{\beta}}{2\sqrt2}\,, \quad c_4=-\fft{{\rm
i}\sqrt{\beta}}{2\sqrt2}\,,\quad c_5=-\fft{{\rm
i}\sqrt{\beta}}4\,,\quad c_6=\fft{\beta\,a}{4\sqrt2}\,.\label{kkec}
\end{eqnarray}

\section{Pseudo-supersymmetrizing scalar/gravity}

As discussed in the introduction, our ultimate goal of this paper is
to construct gauged Kaluza-Klein pseudo-supergravity. The gauging
process will generate a scalar potential. In this section, as a
warmup exercise, we consider the pseudo-supersymmetrization of the
scalar/gravity system.  The bosonic Lagrangian is given by
\begin{equation}
e^{-1}{\cal L}_{{\rm scalar},B} = R - \ft12 (\partial\phi)^2 -
V(\phi)\,.
\end{equation}
The potential $V$ can be expressed in terms of a superpotential $W$, {\it via}
\begin{equation}
V=W'^2 - \fft{D-1}{2(D-2)} W^2\,,\label{vwrelation}
\end{equation}
where a prime denotes a derivative with respect to $\phi$.  Killing
spinor equations for this system in the context of domain wall
solution were given in \cite{fnss}. We propose the ansatz for the
fermionic extension
\begin{eqnarray}
e^{-1} {\cal L}_{{\rm scalar},F} &=& s^{ij}\Big[\ft12 \bar
\psi^i_{\mu} \Gamma^{\mu\nu\rho} D_\nu \psi^j_{\rho} + \ft12 \bar
\lambda^i {\slashed D} \lambda^j+ e_1\, \bar \psi^i_\mu \Gamma^\nu
\Gamma^\mu \lambda^j \partial_\nu \phi\Big]\cr 
&&+ u^{ij}\Big[e_6\, \bar\psi^i_\mu \Gamma^{\mu\nu} \psi^j_\nu\, W+
e_7\, \bar\psi^{i\mu}\Gamma^\mu \lambda^j\,W' + \bar \lambda^{i}
\lambda^{j}\,(e_8 W'' + e_9 W)\Big]\,.
\end{eqnarray}
Note that $e_1$ is given by (\ref{kkec}) and $e_6,\ldots, e_9$ are to
be determined.  The ansatz for the pseudo-supersymmetric transformation rules is given by
\begin{eqnarray}
\delta\psi^i_\mu &=& D_\mu \epsilon^i+ \ft{1}{2\sqrt2\,(D-2)}
b_1\,u^{ij}s^{kj}\,W \Gamma_\mu\epsilon^k\,,\cr 
\delta \lambda^i &=& c_2 \Big[\Gamma^\mu
\partial_\mu \phi\epsilon^i -\sqrt2\,b_2\,u^{ij}s^{kj}\,
W'\epsilon^k\Big]\,,\cr 
\delta e^a_\mu &=& \ft14 s^{ij}\,\bar \psi^i_\mu \Gamma^a
\epsilon^j\,,\cr 
\delta\phi &=& c_4\, s^{ij}\,\bar \lambda^i
\epsilon^j\,.\label{scalartrans}
\end{eqnarray}
The coefficients $c_2$ and $c_4$ are given by (\ref{kkec}) and
coefficients $b_1$ and $b_2$ are to be determined.  It is worth
mentioning that $s^{ij}$ and $u^{ij}$, which can be found in Table 1
in the appendix, can be the same only for cases with $\beta =-1$. We find that
the requirement for pseudo-supersymmetry implies that
\begin{equation}
e_6=-\fft{{\rm i}\sqrt{\beta}}{4\sqrt2}\,,\quad e_7=-\fft12\,,\quad
e_8 =\fft{{\rm i}\sqrt{\beta}}{\sqrt2}\,,\quad e_9 =- \fft{{\rm
i}\sqrt{\beta}}{4\sqrt2}\,,\quad b_1=b_2={\rm i}\sqrt{\beta}\,.
\end{equation}

\section{Kaluza-Klein pseudo-supergravity with a scalar potential}

We now combine the results of section 2 and section 3 together. Keep
in mind that in section 3, the superpotential $W$ can be an
arbitrary function of $\phi$.  We shall examine the restriction on
$W$ in Kaluza-Klein pseudo-supergravity. The
full Lagrangian is given by
\begin{eqnarray}
&&e^{-1}{\cal L}_{\rm KK,pot} = R - \ft12(\partial \phi)^2 - \ft14
e^{a\phi} F^2_\2 - V\cr 
&&\qquad+ s^{ij}\Big[\ft12 \bar \psi^i_{\mu} \Gamma^{\mu\nu\rho}
D_\nu \psi^j_{\rho} + \ft12 \bar \lambda^i {\slashed D} \lambda^j+
e_1\, \bar \psi^i_\mu \Gamma^\nu \Gamma^{j\mu} \lambda
\partial_\nu \phi\Big]\cr 
&&\qquad+ t^{ij}\Big[e_2\, \bar\psi^i_\mu \Gamma^{\mu\nu\rho\sigma}
\psi^j_\sigma + e_3\, \bar\psi^{i\nu}\psi^{j\rho} + e_4\, \bar
\psi^i_\mu \Gamma^{\nu\rho} \Gamma^\mu\lambda^j + e_5\, \bar
\lambda^i \Gamma^{\nu\rho}\lambda^j\Big] e^{\fft12a\phi}
F_{\nu\rho}\cr 
&&\qquad + u^{ij}\Big[e_6\, \bar\psi^i_\mu \Gamma^{\mu\nu}
\psi^j_\nu\, W+ e_7\, \bar\psi^{i\mu}\Gamma^\mu \lambda^j\,W' + \bar
\lambda^{i} \lambda^{j}\,(e_8 W'' + e_9 W)\Big]\,.
\end{eqnarray}
The pseudo-supersymmetric transformation rules are given by
\begin{eqnarray}
\delta\psi_\mu^i &=& D_\mu \epsilon^i  + \ft{{\rm i}}{8(D-2)}\, c_1
t^{ij}s^{kj}\Big(\Gamma_\mu \Gamma^{\nu\rho} -2(D-2)
\delta_{\mu}^{\nu} \Gamma^{\rho}\Big) e^{\fft12 a\phi} F_{\nu\rho}
\epsilon^k\cr 
&&\qquad\qquad+ \ft{1}{2\sqrt2\,(D-2)} b_1\,u^{ij}s^{kj}\,W
\Gamma_\mu\epsilon^k\,, \cr 
\delta \lambda^i &=& c_2 \Big[\Gamma^\mu \partial_\mu\phi\epsilon^i
+ \ft{{\rm i}}{4} c_3\,a\, t^{ij}s^{kj}\,\Gamma^{\mu\nu}
e^{\fft12a\phi}F_{\mu\nu} \epsilon^k  -\sqrt2\,b_2\,u^{ij}s^{kj}\,
W'\epsilon^k \Big]\,,\cr 
\delta e^a_\mu &=& \ft14 s^{ij}\,\bar \psi_\mu^i \Gamma^a
\epsilon^j\,,\cr 
\delta\phi &=& c_4\, s^{ij}\, \bar \lambda^i \epsilon^j\,,\cr 
\delta A_\mu &=& e^{-\fft12a\phi}\, t^{ij}\Big[ c_5\, \bar\psi_\mu^i \epsilon^j
 + c_6\, \bar\lambda^i \Gamma_\mu \epsilon^j
\Big]\,,\label{kkpottrans}
\end{eqnarray}
We find that the variation of the Lagrangian, up to quadratic order
in fermions, yields
\begin{eqnarray}
\delta {\cal L} &=&  -\fft{a\sqrt{\beta}}{8}\gamma(W'' +
\fft{\sqrt2}{\sqrt{(D-1)(D-2)}} W' - \fft{D-3}{2(D-2)} W) \bar
\lambda^i \Gamma^{\mu\nu} \epsilon^j \varepsilon^{ij} e^{\fft12
a\phi} F_{\mu\nu}\cr 
&& - \fft{1}{8\sqrt2} \gamma \Big(  a\,W' - \fft{D-3}{D-2} W\Big)
e^{\fft12a\phi}\bar \psi^i_{\mu}
\Gamma^{\mu\nu\rho} \epsilon^j \varepsilon^{ij} e^{\fft12 a\phi}
F_{\nu\rho}\,,
\end{eqnarray}
where $\gamma=\pm 1$ is given by (\ref{gammavalue}) in the appendix.
The vanishing of the above variation requires that
\begin{equation}
W=m \,e^{\fft{D-3}{\sqrt{2(D-1)(D-2)}}\,\phi} \qquad
\Longrightarrow\qquad V=-\fft{2m^2}{D-1}
e^{\fft{2(D-3)}{\sqrt{2(D-1)(D-2)}}\,\phi}\,,
\end{equation}
where $m$ is a free parameter.  Thus we see that in Kaluza-Klein
pseudo-supergravity, the scalar potential cannot be arbitrary, but a
specific single exponential structure. Note that $(s,t,u)$ cannot be all the same in this case.  In dimensions where Majorana spinors are allowed, two copies are needed with $\varepsilon^{ij}$ and $\delta^{ij}$ bilinear structures.  The global symmetry is $U(1)$, which is a subgroup of $SU(2)$.  In dimensions where symplectic Majorana is necessary, there is also additional $\delta^{ij}$ bilinear structure.  The global symmetry is broken down from $Sp(2)$ to $U(1)$.  In the next section, we consider gauging the $U(1)$ global symmetry.

\section{Gauged Kaluza-Klein pseudo-supergravity }

   In the previous sections, we consider Kaluza-Klein
pseudo-supergravities where the fermions are all neutral under the
Kaluza-Klein vector $A_\1$.  In this section, we gauge the theory by
considering that all the fermions are charged under $A_\1$. As we
shall see presently, this turns the global symmetry of the $U(1)$
rotation of the fermions into a local symmetry. The charged covariant
derivative on fermions is given by
\begin{equation}
D_\mu\xi^i\rightarrow {\cal D}(A) \xi^i = D_\mu \xi^i + b\, A_\mu\,
s^{li}u^{lk}s^{mk}t^{mj}\xi^j =D_\mu \xi^i - \beta\gamma\,b\,
A_\mu\, \varepsilon^{ij}\xi^j \,,
\end{equation}
where $\gamma=\pm 1$ is given by (\ref{gammavalue}), and the charge
parameter $b$ is a constant to be determined.  Note that here
$\xi^i$ represents both $\psi_\mu^i$ and $\lambda^i$.  The full Lagrangian for gauged pseudo-supergravity is now given by
\begin{equation}
{\cal L}_{\rm KK,gauged} = {\cal L}_{\rm KK,pot}(D\rightarrow {\cal
D}(A))\,.
\end{equation}
The pseudo-supersymmetric transformation rules also take the same
form as (\ref{kkpottrans}), but with the covariant derivative $D$ on
the spinors replaced by ${\cal D}(A)$.

  We find that the variation of the Lagrangian ${\cal L}_{\rm KK,gauged}$
leads to the following
\begin{eqnarray}
\delta {\cal L} &=& -\fft{a\sqrt{\beta}}{8}\gamma(W'' +
\fft{\sqrt2}{\sqrt{(D-1)(D-2)}} W' - \fft{D-3}{2(D-2)} W) e^{\fft12
a\phi} \bar \lambda^i \Gamma^{\mu\nu} \epsilon^j \varepsilon^{ij}
F_{\mu\nu}\cr 
&&- \fft{1}{8\sqrt2} \gamma \Big( ( a\,W' - \fft{D-3}{D-2} W)
e^{\fft12a\phi} + \ft12\beta\, b\Big)\bar \psi^i_{\mu}
\Gamma^{\mu\nu\rho} \epsilon^j \varepsilon^{ij} F_{\nu\rho}\,.
\end{eqnarray}
Thus we have
\begin{equation}
W=\fft{g}{\sqrt2}\Big((D-3) e^{-\fft{D-1}{\sqrt{2(D-1)(D-2)}}\,
\phi} + (D-1) e^{\fft{D-3}{\sqrt{2(D-1)(D-2)}}\,
\phi}\Big)\,.\label{suppot}
\end{equation}
The corresponding scalar potential is
\begin{equation}
V=-g^2 (D-1) \Big( (D-3) e^{-\sqrt{\fft{2}{(D-1)(D-2)}}\,\phi} +
e^{\fft{\sqrt2\,(D-3)}{\sqrt{(D-1)(D-2)}}\,\phi}\Big)\,.
\label{f2pot}
\end{equation}
Note that the potential has a maximum and we have chosen the
parameters so that it occurs at $\phi=0$, with $V(0)=-(D-1)(D-2)
g^2$.  The charging parameter $b$ is given by
\begin{equation}
b=-\fft{\beta} 4(D-3)g\,.
\end{equation}
Note that in $D=3$, appropriate scaling limit should be performed to
obtain non-vanishing results. Under the local gauge transformation,
\begin{equation}
A_\1\rightarrow A_\1+ d\Lambda\,,
\end{equation}
the fermions transform as follows
\begin{eqnarray}
\left(\begin{array}{c}
\xi^1 \\
\xi^2 \\
\end{array}
\right) \rightarrow \exp\left[\theta \left(
\begin{array}{cc}
0 & 1\\
-1 & 0\\
\end{array}
\right)\right] \left(
             \begin{array}{c}
               \xi^1 \\
               \xi^2 \\
             \end{array}
           \right)
=\left(
  \begin{array}{cc}
    \cos\theta & \sin\theta \\
    -\sin\theta  & \cos\theta \\
  \end{array}
\right) \left(
             \begin{array}{c}
               \xi^1 \\
               \xi^2 \\
             \end{array}
           \right)\,,\label{so2}
\end{eqnarray}
or equivalently
\begin{eqnarray}
\xi^1+{\rm i}\,\xi^2\rightarrow e^{-{\rm i}\theta}(\xi^1+{\rm
i}\,\xi^2)\,,
 \end{eqnarray}
where
\begin{equation}
\theta = \fft{\gamma}{4}(D-3)g\Lambda\,.\label{thetavalue}
\end{equation}
Note that the $U(1)\sim SO(2)$ rotation (\ref{so2}) leaves both the structures
$\delta^{ij}$ and $\varepsilon^{ij}$ invariant, and hence the
Lagrangian is invariant under the gauge symmetry.  By gauging,
we see that the global symmetry of the $U(1)$ rotation of the
fermions of the ungauged theory becomes the local symmetry, with the original constant $\theta$ now relating to the gauge parameter $\Lambda$ by (\ref{thetavalue}).

    Thus we obtain the pseudo-supersymmetrization of the bosonic
Lagrangian (\ref{fullbosonlag}).  For $D\ge 4$, the theory cannot be
fully supersymmetrized.  However, it can be embedded in maximally
gauged supergravities in $D=4,5,6$ and 7. The embedding in $D=4,5$ and 7 can be understood
as follows.  The gauged group of maximally gauged supergravities in
these dimensions are $SO(8)$, $SO(6)$ and $SO(5)$ respectively,
which admit $U(1)^N$ truncations with $N=4,3,2$.  The Kaluza-Klein
vector in our theory is the one of the $N$ vectors of the truncated
theories.  (See, for example, \cite{tenauthor}.)
The $D=6$ example can be embedded \cite{clpbubble} in six-dimensional
gauged supergravity \cite{romans} with a vector multiplet, which has
an origin \cite{clp6to10,clpbubble} in massive type IIA supergravity
\cite{romans10}.

\section{Conclusions}

In this paper, we construct gauged Kaluza-Klein AdS
pseudo-supergravity in diverse dimensions.  By pseudo-supergravity, we mean that the full Lagrangian is invariant under pseudo-supersymmetric transformation rules up to quadratic fermion order. We start with
pseudo-supersymmetrizing the $D$-dimensional Kaluza-Klein theory that is
the $S^1$ reduction of pure gravity in $D+1$ dimensions. We then
consider pseudo-supersymmetrizing the scalar/gravity system.
Combining the two results, we obtain the pseudo-supersymmetric
Kaluza-Klein theory with a single-exponential scalar potential.  By
requiring that the fermions are all charged under the Kaluza-Klein
vector, we obtain gauged Kaluza-Klein pseudo-supergravity.  In dimensions where there can be Majorana spinors, two copies are needed, with bilinear structures of either $\delta^{ij}$ or $\varepsilon^{ij}$, which has a $U(1)$ global symmetry. In dimensions where there can be symplectic-Majorana, the global symmetry is also $U(1)$, which is a subgroup of $Sp(2)$.  The effect of the gauging is that the $U(1)$ symmetry becomes a local one, associated with the Kaluza-Klein vector. The scalar potential now involves two exponential terms and it has a maximum, giving rise to the AdS spacetime as its vacuum solution.

    The success of our construction, together with the previous
example of pseudo-super- symmetrizing of the bosonic string
\cite{lpwpsugr}, suggests that when a bosonic system admits
consistent Killing spinor equations, it can always be
pseudo-supersymmetrized. In dimensions when the fermion and boson
degrees of freedom happen to match, full supersymmetry may be
realized.

      The highest dimension in gauged supergravities with AdS vacua
is $D=7$. In our gauged pseudo-supergravities, it can be arbitrary.
Solutions of charged rotating \cite{wu} and static black holes
\cite{llw} were previously obtained.  These solutions provide
interesting higher dimensional backgrounds to test the AdS/CFT
correspondence. The pseudo-supersymmetry that our theories possess
makes them superior to an ad hoc concocted AdS theory.  The existence of
consistent Killing spinor equations implies that we can in principle
derive the complete set of solutions that preserves
pseudo-supersymmetry. These solutions are as good as BPS solutions
in gauged supergravities in testing the AdS/CFT correspondence.

       Finally, we would like to emphasize that the possible
bosonic theories that can be pseudo-supersymmetrized are rather
limited.  It was shown that Einstein gravity coupled to an $n$-form
field strength cannot in general pseudo-supersymmetrized, unless it
happens to be part of a supergravity theory \cite{lw0}. The known
non-trivial examples of pseudo-supergravities constructed so far,
besides our examples here, are the pseudo-supergravity extensions of
the bosonic string \cite{lpwpsugr}.  It is of great interest to
investigate whether there exists a classification scheme of
pseudo-supergravities, as in the case of supergravities.

\section*{Acknowledgement}

We are grateful to Yi Pang and Chris Pope for useful discussions.
H.~Liu is grateful to KITPC, Beijing, for hospitality during the
course of this work, and is supported in part by the National
Science Foundation of China (10425525,10875103), National Basic
Research Program of China (2010CB833000) and Zhejiang University
Group Funding (2009QNA3015). Z.L.~Wang is supported by the National
Research Foundation of Korea(NRF) with grant number 2010-0013526.

\appendix
\section{Summary of the results}

The field content of the gauged Kaluza-Klein AdS pseudo-supergravity we have constructed consists of the metric, the dilaton $\phi$ and the Kaluza-Klein vector $A_\1$, together with the pseudo-supersymmetric partners, pseudo-gravitino and pseudo-dilatino $(\psi_\mu^i, \lambda^i)$.
The total on-shell degrees of freedom of bosonic fields are $\ft12(D+1)(D-2)$; the ones of the fermionic fields are $(D-2)2^{[\fft{D}{2}]}$.  Thus the theory does not have real supersymmetry, except for $D=3$; the ungauged $D=3$ theory is simply the circle reduction of $D=4$, ${\cal N}=1$ Poincar\'e supergravity. The full Lagrangian in general dimensions is given by
\begin{eqnarray}
&&e^{-1}{\cal L}_{\rm KK, gauged} = R - \ft12(\partial \phi)^2 -
\ft14 e^{a\phi} F^2_\2 - V\cr 
&&\quad+ s^{ij}\Big[\ft12 \bar \psi^i_{\mu} \Gamma^{\mu\nu\rho}
{\cal D}_\nu(A) \psi^j_{\rho} + \ft12 \bar \lambda^i {\slashed {\cal
D}}(A) \lambda^j+ \ft{\rm i}{2\sqrt{2\beta}}\, \bar \psi^i_\mu
\Gamma^\nu \Gamma^{\mu} \lambda^j
\partial_\nu \phi\Big]\cr 
&&\quad+  t^{ij}\Big[\ft{{\rm i} \sqrt{\beta}}{16}\, \bar\psi^i_\mu
\Gamma^{\mu\nu\rho\sigma} \psi^j_\sigma - \ft{{\rm
i}\sqrt{\beta}}{8}\, \bar\psi^{i\nu}\psi^{j\rho} + \ft{a}{8\sqrt2}\,
\bar \psi^i_\mu \Gamma^{\nu\rho} \Gamma^\mu\lambda^j -\ft{{\rm
i}\,D\sqrt{\beta}}{16(D-2)}\, \bar \lambda^i
\Gamma^{\nu\rho}\lambda^j\Big] e^{\fft12a\phi} F_{\nu\rho}\cr 
&&\quad +  u^{ij}\Big[-\ft{{\rm i}\sqrt{\beta}}{4\sqrt2}\,
\bar\psi^i_\mu \Gamma^{\mu\nu} \psi^j_\nu\, W -\ft12\,
\bar\psi^{i}_{\mu}\Gamma^\mu \lambda^j\,W' + {\rm
i}\sqrt{\ft{\beta}2}\,\bar \lambda^{i} \lambda^{j}\,(W'' - \ft14
W)\Big]\,.
\end{eqnarray}
where $F_\2=dA_\1$, $a^2=2(D-1)/(D-2)$ and the charged covariant
derivative is given by
\begin{equation}
{\cal D}(A) \xi^i = D_\mu \xi^i - \beta\gamma\,b\,
A_\mu\, \varepsilon^{ij}\xi^j \,,
\end{equation}
for any fermion $\xi^i$, where $\gamma=\pm 1$, given by (\ref{gammavalue}).  The superpotential $W$ and the potential
$V$ are given by (\ref{suppot}) and (\ref{f2pot}) respectively.
The quantities $(s^{ij},t^{ij},u^{ij})$ take either $\delta^{ij}$ or $\varepsilon^{ij}$, satisfying the following identities
\begin{equation}
s^{ij}s^{ik}=\delta^{jk}\,,\qquad s^{jk}t^{jl}s^{lm}=t^{km}\,,\qquad
s^{jk}t^{jl}=t^{kj}s^{lj}\,,\qquad s^{kj}t^{jl}=t^{kj}s^{jl}\,.
\end{equation}
Note that in the above, $s$ and $t$ can interchange, and each can
interchange with $u$ and the identities still hold.

The pseudo-supersymmetric transformation rules for all the involved
fields are given by
\begin{eqnarray}
\delta\psi^i_\mu &=& \Big[{\mathcal D}_\mu {{} }\epsilon^i  +
\ft{{\rm i}\sqrt{\beta}}{8(D-2)}\, {{} } t^{ij}{{} }
s^{kj}\Big(\Gamma_\mu \Gamma^{\nu\rho} -2(D-2) \delta_{\mu}^{\nu}
\Gamma^{\rho}\Big) e^{\fft12 a\phi} F_{\nu\rho}{{} }\epsilon^k \cr
&&\qquad\quad+ \ft{{\rm i}\sqrt{\beta}}{2\sqrt2\,(D-2)}\,u^{ij}{{} }
s^{kj}\, W \Gamma_\mu{{} }\epsilon^k\Big]\,,\cr 
\delta \lambda^i &=& \ft{{\rm i}\sqrt{\beta}}{2\sqrt2}\Big[
\Gamma^\mu
\partial_\mu\phi\,{{} }\epsilon^i + \ft{{\rm i}\sqrt{\beta}}{4} a\, {{} }
t^{ij}{{} } s^{kj}\,e^{\fft12 a\phi}F_{\mu\nu}\,\Gamma^{\mu\nu}{{}
}\epsilon^k -{\rm i}\sqrt{2\beta}\, {{} } u^{ij}{{} } s^{kj}\, W'{{}
}\epsilon^k\Big]\,,\cr 
\delta e^a_\mu &=& \ft14{{} } s^{ij} \,\bar \psi^i_\mu \Gamma^a {{}
}\epsilon^j\,,\qquad \hbox{so}\qquad \delta g_{\mu\nu} = \ft12 {{} }
s^{ij}\bar\psi^i_{(\mu} \Gamma_{\nu)}{{} } \epsilon^j\,,\cr 
\delta\phi &=& -\ft{{\rm i}\sqrt{\beta}}{2\sqrt2}\, {{} }
s^{ij}\,\bar\lambda^i {{} }\epsilon^j \,,\cr 
\delta A_\mu &=& e^{-\fft12a\phi}\, {{} } t^{ij}\,\Big[-\ft{{\rm
i}\sqrt{\beta}}{4}\, \bar\psi^i_\mu {{} }\epsilon^j  +
\ft{a\,\beta}{4\sqrt2}\, \bar \lambda^i \Gamma_\mu {{} }\epsilon^j
\Big]\,.
\end{eqnarray}

To verify that the Lagrangian is indeed invariant under the transformation rules, up to the quadratic fermion order, it is useful to derive first the following projected integrability conditions, given by
\begin{eqnarray}
&&s^{ji}\Gamma^{\mu\nu\rho}{\cal D}_{\nu}\delta\psi^j_\rho\cr 
&=& \ft12 \Gamma^{\nu}\left[R{}{}^{\mu}{}_{\nu}-
\ft12\partial^\mu\phi\,\partial_\nu\phi - \ft12F^{2\mu}{}_{\nu}
-\ft12\delta^{\mu}{}_{\nu} \left(R-\ft12(\partial\phi)^2 -
\ft14F^2-V\right)\right]\epsilon^i \cr 
&&+\ft{{\rm i}\sqrt{\beta}}{8}s^{ji}t^{jk}s^{mk} \Big(2g^{\mu_1\nu}
g^{\mu\mu_2}-\Gamma^{\mu\nu\mu_1\mu_2}\Big)e^{\fft12 a\phi}
F_{\mu_1\mu_2}\delta\psi_{\nu}^m +\ft{{\rm
i}\sqrt{\beta}}{2\sqrt2}s^{ji}u^{jk}s^{mk}
\Gamma^{\mu\nu}W\delta\psi_{\nu}^m \cr 
&&-\ft{\rm i}{2\sqrt{2\beta}} s^{ji}\nabla_{\nu}\phi\,
\Gamma^{\nu}\Gamma^{\mu}\delta\lambda^j-\ft{\sqrt2}{16}
s^{ji}t^{jk}s^{mk}a\,e^{\fft12a\phi}F_{\mu_1\mu_2}
\Gamma^{\mu_1\mu_2}\Gamma^{\mu}\delta\lambda^m \cr 
&& +\ft12 s^{ji} u^{jl}s^{kl}\,W'\Gamma^{\mu}\delta\lambda^l
+\ft{{\rm i}\sqrt{\beta}}{8}s^{ji}t^{jk} \Big(2e^{-\fft12
a\phi}\nabla_{\nu}(e^{ a\phi} F^{\nu\mu})-e^{\fft12
a\phi}\Gamma^{\mu\nu\mu_1\mu_2}\nabla_{\nu}
F_{\mu_1\mu_2}\Big)\epsilon^k  \cr
&&- \ft{1}{8\sqrt2}\gamma \varepsilon^{ij} \left[(\ft{D-3}{D-2}W-a\,W')e^{\fft12
a\phi} -\ft{\beta}2\,b \right]\Gamma^{\mu\nu\rho} F_{\nu\rho}
\epsilon^j\,,\label{interpsi}
\end{eqnarray}
and
\begin{eqnarray}
&&s^{ji}\Gamma^{\mu}{\cal D}_{\mu}\delta\lambda^j\cr 
&=& \ft{{\rm i}\sqrt{\beta}}{2\sqrt2} s^{ji}\,\Gamma^{\mu}\Gamma^\nu
\partial_\nu\phi\,\delta\psi^j_{\mu} -\ft{\beta\,a}{8\sqrt2}
s^{ji}t^{jk}s^{lk}e^{\fft12a\phi}F_{\mu_1\mu_2}
\Gamma^{\mu}\Gamma^{\mu_1\mu_2} \delta\psi_\mu^l +\ft12\beta
s^{ji}u^{jk}s^{lk}\,\Gamma^{\mu}W'\,\delta\psi^l_{\mu} \cr
&& +\ft{{\rm i}D\sqrt{\beta}}{8(D-2)} s^{ji} t^{jk}s^{lk}
e^{\fft12a\phi}F_{\mu_1\mu_2}\Gamma^{\mu_1\mu_2}\delta\lambda^l +
\ft{{\rm i}\sqrt{\beta}}{2\sqrt2\,} s^{ji}u^{jk}s^{lk}\,
W\,\delta\lambda^l \cr 
&&-{\rm i}\sqrt{2\beta}\,s^{ji}u^{jk}
s^{lk}\,W''\delta\lambda^l +\ft{{\rm i}\sqrt{\beta}}{2\sqrt2}\left(\nabla^2\phi-\ft{a}{4}
e^{a\phi}F^2-V'\right)\epsilon^i \cr
&&
-\ft{\beta\,a}{8\sqrt2}s^{ji}t^{jk}
\left[e^{\fft12a\phi}\Gamma^{\mu\mu_1\mu_2}\nabla_{\mu}
F_{\mu_1\mu_2}+2e^{-\fft12a\phi}\Gamma^{\mu_2}\nabla^{\mu}
\left(e^{a\phi}F_{\mu\mu_2}\right)\right]\epsilon^k \cr 
&&-\ft{{\rm i}\sqrt{\beta}\,a}{8} \gamma \varepsilon^{ij}\,
e^{\fft12a\phi}F_{\mu_1\mu_2}\Gamma^{\mu_1\mu_2}
\Big[W''+\ft{2}{(D-2)a}W'-\ft{(D-3)}{2\,(D-2)}W  \Big]
\epsilon^m\,.\label{interlam}
\end{eqnarray}
Note that for Killing spinors, we have $\delta\psi_\mu^i=0$ and
$\delta\lambda^i=0$.  Substituting these into the above equations
and we find that what remain are precisely the full set of bosonic
equations of motion attached to various $\Gamma$-matrix structures.

It is worth pointing out that since $s^{ij}=t^{ij}$ occurs only for $\beta=+1$ and $s^{ij}=u^{ij}$ only for $\beta=-1$, the quantities $(s,t,u)$ cannot be all the same in any dimensions.  In dimensions where Majorana spinors are allowed, two copies are necessary with both $\delta^{ij}$ and $\varepsilon^{ij}$ bilinear structures.  When symplectic Majorana is available, the symplectic structure is broken down to include $\delta^{ij}$ structure as well.

     Finally we present the $\Gamma$-matrix and fermion conventions.
We adopt exactly the same convention given in \cite{lpwpsugr}, which
follows the convention of \cite{vanp}.  We present the convention in
Table 1. In addition to the $\Gamma$-matrix symmetries and spinor
representations in diverse dimensions, we also present the
$s^{ij}$, $t^{ij}$ and $u^{ij}$ that appear in the construction. From Table 1, we can derive the following important identities
\begin{equation}
s^{ji}t^{jk} s^{mk}t^{ml}=\beta\, \delta^{il}\,,\qquad
s^{ji}u^{jk}s^{mk}u^{ml}=-\beta\, \delta^{il}\,,\qquad s^{ji}u^{jk}
s^{lk}t^{lm} = \gamma \beta\,\varepsilon^{im}\,,
\end{equation}
where
\begin{equation}
\gamma =\Big\{ \begin{array}{cc} +1\,, & {\rm if~~}
t^{ij}=\delta^{ij} \,,\cr -1\,, & {\rm if~~} t^{ij}=\varepsilon^{ij}
\,.
\end{array}\label{gammavalue}
\end{equation}

\bigskip\bigskip
\centerline{
\begin{tabular}{|c|c|c||c|c||c|c|c|}\hline
 $D$ mod 8 & $C\Gamma^{(0)}$ & $C\Gamma^{(1)}$ & Spinor &$\beta$
& $s^{ij}$ & $t^{ij}$ & $u^{ij}$\\ \hline\hline 
0&S&S&M&$+1$&$\delta^{ij}$&$\delta^{ij}$&$\varepsilon^{ij}$ \\
&S&A&S-M&$-1$&$\varepsilon^{ij}$&$\delta^{ij}$&$\varepsilon^{ij}$\\
\hline 
1&S&S&M&$+1$&$\delta^{ij}$&$\delta^{ij}$&$\varepsilon^{ij}$\\
\hline 
2&S&S&M&$+1$&$\delta^{ij}$&$\delta^{ij}$&$\varepsilon^{ij}$\\
&A&S&M&$-1$&$\delta^{ij}$&$\varepsilon^{ij}$&$\delta^{ij}$\\
\hline 
3&A&S&M&$-1$&$\delta^{ij}$&$\varepsilon^{ij}$&$\delta^{ij}$\\ \hline
4&A&S&M&$-1$&$\delta^{ij}$&$\varepsilon^{ij}$&$\delta^{ij}$\\
&A&A&S-M&$+1$&$\varepsilon^{ij}$&$\varepsilon^{ij}$&$\delta^{ij}$\\
\hline 
5&A&A&S-M&$+1$&$\varepsilon^{ij}$&$\varepsilon^{ij}$&$\delta^{ij}$\\
\hline 
6&A&A&S-M&$+1$&$\varepsilon^{ij}$&$\varepsilon^{ij}$&$\delta^{ij}$\\
&S&A&S-M&$-1$&$\varepsilon^{ij}$&$\delta^{ij}$&$\varepsilon^{ij}$\\
\hline 
7&S&A&S-M&$-1$&$\varepsilon^{ij}$&$\delta^{ij}$&$\varepsilon^{ij}$\\
\hline
\end{tabular}}
\bigskip

\begin{center}
Table 1: $\Gamma$-matrix symmetries,  spinor representations and $(s,t,u)$ in diverse dimensions.
\end{center}


\begin{thebibliography}{99}

\bibitem{cjs} E.~Cremmer, B.~Julia and J.~Scherk,
{\it Supergravity theory in eleven-dimensions,}
  Phys.\ Lett.\  B {\bf 76}, 409 (1978).

\bibitem{fog}
  J.M.~Figueroa-O'Farrill and G.~Papadopoulos,
{\it Maximally supersymmetric solutions of ten-dimensional and
eleven-dimensional supergravities,}
  JHEP {\bf 0303}, 048 (2003)
  [arXiv:hep-th/0211089].

\bibitem{gp}
  J.P.~Gauntlett and S.~Pakis,
{\it The geometry of $D = 11$ Killing spinors,}
  JHEP {\bf 0304}, 039 (2003)
  [arXiv:hep-th/0212008].

\bibitem{lw0}
  H.~L\"u and Z.L.~Wang,
{\it Pseudo-Killing spinors, pseudo-supersymmetric $p$-branes,
bubbling and less-bubbling AdS spaces,} JHEP {\bf 1106}, 113 (2011),
arXiv:1103.0563 [hep-th].

\bibitem{lpw}
  H.~L\"u, C.N.~Pope and Z.L.~Wang,
{\it Pseudo-supersymmetry, consistent sphere reduction and Killing
spinors for the bosonic string,}
  arXiv:1105.6114 [hep-th], to appear in PLB.

\bibitem{lw}
  H.~L\"u and Z.L.~Wang,
{\it Killing spinors for the bosonic string,}
  arXiv:1106.1664 [hep-th].

\bibitem{llw}
  H.~Liu, H.~L\"u and Z.L.~Wang,
{\it Killing spinors for the bosonic string and the Kaluza-Klein
theory with scalar potentials,}
  arXiv:1106.4566 [hep-th].

\bibitem{lpwpsugr} H.~L\"u, C.N.~Pope and Z.L.~Wang,
{\it Pseudo-supergravity extension of the bosonic string,}
  arXiv:1106.5794 [hep-th].

\bibitem{Bvn} E. Bergshoeff, M. de Roo, B. de Wit and P. van Nieuwenhuizen,
{\it Ten-dimensional Maxwell-Einstein supergravity, its currents,
and the issue of its auxiliary fields}, Nucl. Phys. {\bf B195}, 97
(1982).

\bibitem{berderoo} E. Bergsoeff and M. de Roo, {\it Supersymmetric
Chern-Simons terms in ten dimensions}, Phys. Lett. {\bf B218}, 210
(1989).

\bibitem{berderoo2} E.A. Bergshoeff and M. de Roo, {\it The quartic
effective action of the heterotic string and supersymmetry}, Nucl.
Phys. {\bf B328}, 439 (1989).

\bibitem{vanp} A. Van Proeyen,
{\it Tools for supersymmetry}, hep-th/9910030.

\bibitem{fnss}
  D.Z.~Freedman, C.~Nunez, M.~Schnabl and K.~Skenderis,
{\it Fake supergravity and domain wall stability,}
  Phys.\ Rev.\  D {\bf 69}, 104027 (2004)
  [arXiv:hep-th/0312055].

\bibitem{tenauthor}
M.~Cveti\v c, M.J.~Duff, P.~Hoxha, James T.~Liu, H.~L\"u, J.X.~L\"u,
R.~Martinez-Acosta, C.N.~Pope, H.~Sati, T.A.~Tran, {\it Embedding
AdS black holes in ten-dimensions and eleven-dimensions,}
  Nucl.\ Phys.\  B {\bf 558}, 96 (1999)
  [arXiv:hep-th/9903214].

\bibitem{clpbubble}
  Z.W.~Chong, H.~L\"u and C.N.~Pope,
{\it BPS geometries and AdS bubbles,}
  Phys.\ Lett.\  B {\bf 614}, 96 (2005)
  [arXiv:hep-th/0412221].

\bibitem{romans} L.J.~Romans,
{\it The $F_4$ gauged supergravity in six dimensions,}
  Nucl.\ Phys.\  B {\bf 269}, 691 (1986).

\bibitem{clp6to10}
  M.~Cveti\v c, H.~L\"u and C.N.~Pope,
{\it Gauged six-dimensional supergravity from massive type IIA,}
  Phys.\ Rev.\ Lett.\  {\bf 83}, 5226 (1999)
  [arXiv:hep-th/9906221].

\bibitem{romans10}
  L.J.~Romans,
{\it Massive $N=2a$ supergravity in ten-dimensions,}
  Phys.\ Lett.\  B {\bf 169}, 374 (1986).

\bibitem{wu} S.Q. Wu,
{\it General rotating charged Kaluza-Klein AdS black holes in
higher dimensions}, Phys. Rev. D {\bf 83} (2011) 121502 (R).

\end{thebibliography}
\end{document}